\documentclass[a4paper, prd, twocolumn, superscriptaddress]{revtex4-1}

\usepackage{color}
\usepackage{multirow}
\usepackage{amstext}
\usepackage{amssymb}
\usepackage{amsmath}
\usepackage{hyperref}
\usepackage{paralist}
\usepackage{graphicx}
\usepackage{url}

\def\d{{\mathrm{d}}}
\def\beq{\begin{equation}}\def\eeq{\end{equation}}

\def\l{\ell}
\def\lmn{(\ell, m, n)}
\def\lm{(\ell, m)}
\newfont{\cursive}{pzcmi at 9pt}

\newcommand{\gwparams}{\ensuremath{\vec{\vartheta}}}
\newcommand{\likelihood}{p(d|\gwparams,H)}
\newcommand{\prior}{p(\gwparams|H)}
\newcommand{\evidence}{p(d|H)}
\newcommand{\posterior}{p(\gwparams|d,H)}
\newcommand{\BF}{\mathcal{B}_{AB}}
\newcommand{\Ms}{M^{(\text{src})}}

\begin{document}

\title[]{The next decade of black hole spectroscopy}

\author{Miriam Cabero}
\email{mcmuller@princeton.edu}
\affiliation{Department of Physics, Princeton University, Princeton, NJ 08544, USA}

\author{Julian Westerweck}
\affiliation{Max Planck Institute for Gravitational Physics (Albert Einstein Institute), Callinstrasse 38, D-30167 Hannover, Germany}
\affiliation{Leibniz Universit\"at Hannover, Welfengarten 1-A, D-30167 Hannover, Germany}

\author{Collin D. Capano} 
\affiliation{Max Planck Institute for Gravitational Physics (Albert Einstein Institute), Callinstrasse 38, D-30167 Hannover, Germany}
\affiliation{Leibniz Universit\"at Hannover, Welfengarten 1-A, D-30167 Hannover, Germany}

\author{Sumit Kumar}
\affiliation{Max Planck Institute for Gravitational Physics (Albert Einstein Institute), Callinstrasse 38, D-30167 Hannover, Germany}
\affiliation{Leibniz Universit\"at Hannover, Welfengarten 1-A, D-30167 Hannover, Germany}

\author{Alex B. Nielsen}
\affiliation{Max Planck Institute for Gravitational Physics (Albert Einstein Institute), Callinstrasse 38, D-30167 Hannover, Germany}
\affiliation{Leibniz Universit\"at Hannover, Welfengarten 1-A, D-30167 Hannover, Germany}
\affiliation{Department of Mathematics and Physics, University of Stavanger, 4036 Stavanger, Norway}

\author{Badri Krishnan}
\affiliation{Max Planck Institute for Gravitational Physics (Albert Einstein Institute), Callinstrasse 38, D-30167 Hannover, Germany}
\affiliation{Leibniz Universit\"at Hannover, Welfengarten 1-A, D-30167 Hannover, Germany}

\begin{abstract}

  Gravitational wave observations of the ringdown of the remnant black
  hole in a binary black hole coalescence provide a unique opportunity
  of confronting the black hole no-hair theorem in general relativity
  with observational data.  The most robust tests are possible if
  multiple ringdown modes can be observed.  In this paper, using
  state-of-the-art Bayesian inference methods and the most up-to-date
  knowledge of binary black hole population parameters and ringdown
  mode amplitudes, we evaluate the prospects for black hole
  spectroscopy with current and future ground based gravitational wave
  detectors over the next 10 years.  For different population models,
  we estimate the likely number of events for which the subdominant
  mode can be detected and distinguished from the dominant mode.
  We show that black hole spectroscopy could significantly test general
  relativity for events seen by the proposed LIGO Voyager detectors.
  
\end{abstract}

\maketitle

\section{Introduction}

The remnant black hole (BH) formed after the coalescence of two
compact objects emits gravitational radiation while settling down to a
Kerr BH. This stage is known as the ringdown.  Perturbation theory
predicts that, at late enough times, the ringdown consists of a
superposition of exponentially damped sinusoids called quasinormal
modes (QNM) \cite{Vishveshwara:1970zz,Chandrasekhar:1975zza} (see also
\cite{Kokkotas:1999bd,Berti:2009kk}). The QNMs are characterized by a
set of complex frequencies $\Omega_{\l mn}$ labeled by three integers; 
$\l,m$ are angular quantum numbers while $n=0,1,2\ldots$ is the
overtone index. According to the no-hair theorem in standard general
relativity (GR), $\Omega_{\l mn}$ is uniquely defined by the BH mass
and spin.  The measurement of multiple QNMs in a BH ringdown, known as
BH spectroscopy, is crucial for robust observational tests of the
no-hair theorem with gravitational waves based only on the ringdown
signal~\cite{Dreyer:2003bv, PhysRevD.73.064030}.

The excitation of different QNMs depends on the nature of the
perturbation, i.e. on the properties of the binary progenitor
\cite{PhysRevD.85.024018, Borhanian:2019kxt, Hughes:2019zmt,
Apte:2019txp,Lim:2019xrb}.  Thus,
for aligned spin systems, the amplitude of the different modes are
determined by the spins of the initial compact objects and the mass
ratio $q = m_1/m_2 \geq 1$, with $m_{1,2}$ the mass of each object.
The ringdown signature is dominated by the fundamental
$(\l, m) = (2, 2)$ mode~\cite{Flanagan:1997sx}. For non-spinning
binaries with equal masses ($q=1$), odd $\l$ modes vanish and the
loudest subdominant mode is the $(\l, m) = (4, 4)$
mode~\cite{PhysRevD.85.024018, Borhanian:2019kxt}. As the mass ratio
increases, the $(\l, m) = (3, 3)$ mode becomes the loudest subdominant
mode, reaching amplitude ratios as large as
$A_{330}/A_{220} \simeq 0.3$.  Hence, coalescences of two unequal-mass
BHs or neutron-star black-hole binaries (NSBH) are the most promising sources
for measurability of subdominant modes in the ringdown.  For still
higher mass ratios, the relative amplitude of the modes can also tell
us about the alignment of the orbit relative to the BH spin
during the inspiral phase
\cite{Hughes:2019zmt,Apte:2019txp,Lim:2019xrb}.

Two main conditions are necessary to test the no-hair
theorem: (i) the detectability of at least two modes, and (ii) the
resolvability of the frequencies and/or damping times of each
mode. Theoretical estimates of the necessary ringdown 
signal-to-noise ratio (SNR) for each of these conditions can be found in the
literature~\cite{PhysRevD.73.064030, PhysRevD.76.104044}. These
studies have predicted that Advanced LIGO should observe several
ringdown events at design sensitivity, but will not be able to detect
subdominant modes from the coalescence of stellar-mass BBH for BH
spectroscopy~\cite{Berti:2016lat, Baibhav:2018rfk}.  In this paper we
revisit the prospects for accurate BH spectroscopy with the next
decade of LIGO detectors. In general, asymmetric binaries are more
likely to produce higher amplitudes for the subdominant ringdown
modes.  However, based on the gravitational-wave observations to 
date, more asymmetric systems are also likely to be much fewer in 
number~\cite{LIGOScientific:2018jsj} (although recent public alerts 
from the third observing run of Advanced LIGO and Virgo suggest
possible detections of NSBH~\cite{NSBH1, NSBH2}).  In addition, the
orientation of a source relative to the detectors also has an
important effect on the observed amplitudes. Systems where the angular
momentum is aligned with the line-of-sight to the source are more
luminous, but these orientations are not favorable for observing the
subdominant modes.  Taking all these effects into account, along with
the most up-to-date estimates of the ringdown mode
amplitudes~\cite{Borhanian:2019kxt} and state-of-the-art gravitational
wave parameter estimation techniques~\cite{pycbc_inference,
  alex_nitz_2019_3265452}, we show that black hole spectroscopy can
provide non-trivial limits on general relativity with the LIGO Voyager
detector.

At least 10 binary black-hole (BBH) coalescences have been observed in
the first two observing runs of Advanced LIGO and
Virgo~\cite{GWTC-1,Nitz:2019hdf,Nitz:2018imz,Antelis:2018smo,Zackay:2019btq,Venumadhav:2019lyq}.
The loudest BBH event is still the first detection,
GW150914~\cite{GW150914}, with a ringdown signal-to-noise ratio (SNR)
$\rho \simeq 8.5$~\cite{TheLIGOScientific:2016src} at 3 ms after
merger.  This event has not provided significant evidence for the
presence of measurable subdominant modes with
$\l \neq 2$~\cite{Carullo:2019flw}.  However, recent work suggests
that the inclusion of higher overtones of the dominant $\l = 2$ mode
allows for the modeling of the ringdown immediately after the merger,
hence obtaining higher SNR in ringdown
signatures~\cite{Giesler:2019uxc}.  The analysis of the GW150914
ringdown using the fundamental mode and its first overtone provides
the first constraints to date of deviations of the no-hair theorem
using two QNMs~\cite{Isi:2019aib}.
Here we use the Bayesian inference and model selection
frameworks~\cite{Kass:1995loi} on simulated BBH populations to establish 
the measurability and accurate resolvability of two ringdown QNMs
over the next decade, providing rate estimates for constraining the 
no-hair theorem to within $\pm 20\%$ at the $90\%$ credible level.
We restrict ourselves to the resolvability of subdominant QNMs ($\l \neq 2$) 
for two reasons:
\begin{inparaenum}[(1)] 
\item the excitation amplitudes of overtones on the general parameter space of
the binary's properties are not yet well-understood and we lack predictions to
model ringdown signatures that include overtones for a large population of BBH
mergers, and 
\item the frequencies of the overtones are very similar to each other, hence
accurate resolvability of an overtone is more challenging than of a subdominant
mode.  
\end{inparaenum} 

This manuscript is organized as follows. Section~\ref{sec:bayesian} introduces
the Bayesian inference and model selection frameworks, as well as the ringdown
model used.  Section~\ref{sec:population} describes the details on the BBH
population considered.  In Section~\ref{sec:results} we report the rates on
measurable subdominant modes and prospects for resolvability of the necessary
parameters to perform tests of the no-hair theorem. Finally, we conclude our
findings in Sec.~\ref{sec:conclusions}.

\section{Bayesian framework}
\label{sec:bayesian}

We use Bayesian methods to infer the properties of the remnant BH from
our data, $d(t)$, and to determine the presence of a measurable
subdominant mode in the ringdown signature.  Given a model hypothesis
of the ringdown signal, $H$, parametrized by the source properties,
\gwparams, Bayes' theorem defines the posterior probability
distribution: \beq \posterior = \prior \frac{\likelihood}{\evidence}
\, , \eeq where $\prior$ is the prior knowledge based on astrophysical
populations or theoretical models, the likelihood $\likelihood$ is the
conditional probability of observing the data $d(t)$ given the model
$H$ with parameters $\gwparams$, and the evidence $\evidence$ is a
normalization constant that only depends on the data and the chosen
model.  Calculating the evidence requires marginalization over the
entire parameter space, which can become computationally
challenging. While this computation can be avoided for Bayesian
parameter estimation, model selection between two competing models
requires accurate estimates of the evidence.

In Bayesian model selection, the Bayes factor weighs the evidence provided
by the data in support of one model versus 
another~\cite{Kass:1995loi, annis2019thermodynamic}:
\beq
\BF = \frac{p(d|H_A)}{p(d|H_B)} \, .
\eeq
The larger $\BF$ is, the stronger the data supports hypothesis $H_A$ over
$H_B$.  Following the nomenclature of~\cite{Kass:1995loi}, a Bayes factor
$>3.2$ indicates ``substantial" support for $H_A$ over $H_B$; $\BF > 10$
indicates ``strong" support, while $\BF > 100$ is ``decisive". A Bayes factor
between $1/3.2$ and $3.2$ is ``not worth mentioning"; i.e., the data is
inconclusive as to whether $H_A$ or $H_B$ is favored.

\subsection{The likelihood function}

For a GW detector network with uncorrelated stationary Gaussian noise,
the likelihood is given by
\beq
\likelihood \propto \exp \left[ -\frac{1}{2} \sum_{a=1}^N \langle d_a - h_a(\gwparams), d_a - h_a(\gwparams) \rangle \right] \, ,
\eeq
where $N$ is the number of detectors, $d_a$ is the data for each detector,
and $h_a(\gwparams)$ is the waveform model evaluated for a set of parameters $\gwparams$ 
as observed by detector $a$. The noise-weighted inner product is defined as
\beq \label{eq:inner_product}
\langle x, y \rangle = 4 \Re \int_0^\infty \frac{\tilde{x}^*(f) \tilde{y}(f)}{S_n(f)} \d f \, ,
\eeq
with $S_n(f)$ being the one-sided power spectral density (PSD) of the detector's noise, 
$\tilde{x}(f)$ the Fourier transform of $x(t)$, and $^*$ indicating the complex conjugate.

In this paper we use the {\texttt{PyCBC Inference}}~\cite{pycbc_inference,
alex_nitz_2019_3265452} toolkit to compute the likelihood function and estimate
posterior probability distributions. Accurate marginalization for evidence
estimation is achieved using the nested sampling algorithm
{\texttt{cpnest}}~\cite{john_veitch_2017_835874}.

\subsection{The ringdown model}
\label{sec:ringdown}

The strain $h(t)$ produced by a gravitational wave at the detector is
given by 
\beq \label{eq:h_det} 
h(t) = F_+(\alpha, \delta, \Psi) h_+(t) + F_\times(\alpha, \delta, \Psi) h_\times(t) \, , 
\eeq 
where $F_{+, \times}$ are the antenna pattern functions determined 
by the relative orientation between the detector frame and the wave
frame~\cite{300years}, i.e. the sky location of the source (right
ascension $\alpha$ and declination $\delta$ in a geocentric coordinate
system) and the polarization angle $\Psi$ that defines the relative
orientation of the wave frame with the geocentric coordinate system.
For short transient signals, these orientation angles (and hence
$F_{+, \times}$) are assumed to be time independent.  For future
generation of observatories with improved low frequency sensitivity,
it might become necessary to account for the time dependence of
$F_{+,\times}$. However, the ringdown itself will be short enough
that for our purposes we do not need to consider this effect here.

The ringdown signal of a Kerr BH consists of a sum of exponentially
damped sinusoids: 
\beq 
h_+ + i h_\times = \frac{M}{D_L} \sum_{\l, m, n} {}_{-2}S_{\l m} (\iota, \varphi) 
A_{\l mn} e^{i(\Omega_{\l mn}t +  \phi_{\l mn})} \, , 
\eeq 
where $M$ is the mass of the BH in the detector frame and $D_L$ 
is the luminosity distance to the source.
The functions ${}_{-2}S_{\l m} (\iota, \varphi)$ are the spin-weighted
spheroidal harmonics, which depend on the inclination angle $\iota$
between the BH spin and the line-of-sight from the observer to the
source, and the azimuth angle $\varphi$ between the BH and the
observer. The complex QNM frequencies $\Omega_{\l mn}$, determined
from the Teukolsky equation \cite{Teukolsky:1972my,Leaver:1985ax},
define the frequency and damping time of the damped sinusoid,
$\Omega_{\l mn} = \omega_{\l mn} + i/\tau_{\l mn}$.  The amplitudes
$A_{\l mn}$ and $\phi_{\l mn}$ depend on the initial perturbation and
take different values for different $\lmn$ modes.
Henceforth, we restrict ourselves to the $n=0$ overtone and drop the 
overtone index $n$ for simplicity.

Assuming that the ringdown begins at $t=0$, the two gravitational-wave polarizations are given by 
\begin{align} \label{eq:ringdown_polarizations}
h_+(t) &= \frac{M}{D_L} \sum_{\ell,m} {}_{-2}Y_{\ell m}^+(\iota) A_{\ell m} e^{-t/\tau_{\ell m}} \cos(\omega_{\ell m} t + \phi_{\ell m}) \, , \nonumber \\ 
h_\times(t) &= \frac{M}{D_L} \sum_{\ell,m} {}_{-2}Y_{\ell m}^\times(\iota) A_{\ell m} e^{-t/\tau_{\ell m}} \sin(\omega_{\ell m} t +\phi_{\ell m}) \, ,
\end{align}
where we have approximated the spheroidal harmonics ${}_{-2} S_{\ell m}$ by 
spin-weighted spherical harmonics ${}_{-2} Y_{\ell m}$~\cite{PhysRevD.73.024013, PhysRevD.76.104044}: 
\begin{align}
{}_{-2}Y_{\ell m}^+(\iota) &= {}_{-2}Y_{\ell m}(\iota, 0) + (-1)^\ell {}_{-2}Y_{\ell \text{-}m}(\iota, 0) \, , \nonumber \\ 
{}_{-2}Y_{\ell m}^\times(\iota) &= {}_{-2}Y_{\ell m}(\iota, 0) -(-1)^\ell {}_{-2}Y_{\ell \text{-}m}(\iota, 0) \, .  
\end{align}

The ringdown analysis in this paper follows the methods developed
in~\cite{Cabero:thesis, 2018PhRvD..97l4069C}. 
We use two different waveform models,
\begin{inparaenum}[(i)]
a \item {\it{Kerr model}} where we assume the remnant object to be a Kerr BH, hence
the ringdown QNM frequencies $\Omega_{\l mn}$ are uniquely determined by the mass $M$ 
and the spin $\chi$ of the BH, and \label{Kerr}
an \item {\it{agnostic model}} where we assume the nature of the remnant object to be unknown,
hence the ringdown is parameterized by each individual QNM frequency $\Omega_{\l mn}$ 
and we drop the factor $M/D_L$ in Eq.~\eqref{eq:ringdown_polarizations}. \label{agnostic}
\end{inparaenum}
The Kerr model~(\ref{Kerr}) is our starting point for determining the measurability 
of a subdominant mode. Resolvability of the subdominant mode for testing the no-hair 
theorem is determined using the agnostic model~(\ref{agnostic}).

\section{Populations}
\label{sec:population}

We construct populations of candidate BBH ringdown signals based on the
observational population model B of~\cite{LIGOScientific:2018jsj}
(we ignore NSBH mergers here because population models including 
NSBH are largely uncertain). 
The component-mass and mass-ratio distributions follow power laws with 
exponents $-\alpha$ and $\beta_q$, respectively (see Eq. (2) in~\cite{LIGOScientific:2018jsj}). 
For the component-mass distribution, the measured median value is $\alpha = 1.6$,
with masses in the range $[5.4,57) M_\odot$ (we use the lowest $m_{min}$ and the largest
$m_{max}$ values, to account for uncertainties in the mass bounds of BHs).
For the mass-ratio distribution we use two different exponent values:
the measured median value $\beta_q = 6.7$, and a uniform distribution $\beta_q = 0$
(which is used in model A of~\cite{LIGOScientific:2018jsj}). 
Mass ratios are restricted to be within the range $[1, 8)$.
We assume the individual BHs to be non-spinning prior to the merger, 
which is consistent with the population of BBHs observed by LIGO/Virgo thus far. 
Sources are distributed uniformly in co-moving volume; we choose a maximum 
luminosity distance, $D_L^{\text{(max)}}$, dependent on the considered detector 
network.  The inclination angle $\iota$ is distributed uniformly 
in $\cos \iota \in [-1, 1)$, and the polarization angle $\psi$ uniformly $\in [0, 2\pi)$. 

The mass and spin of the remnant Kerr BH determine the ringdown
frequencies $\Omega_{\l m}$~\cite{Berti:2005ys}. We obtain an estimate
of the remnant's source frame mass $\Ms$ and dimensionless spin $\chi$ 
using the fitting formulae to numerical relativity~\cite{Tichy:2008du, Hofmann:2016yih} 
implemented in the {\texttt{LALSuite}} software package~\cite{lalsuite}.
The detector frame mass $M$ is given by $M = (1+z)\Ms$, where $z$ is the
redshift calculated from the luminosity distance, $D_L$, assuming a standard 
$\Lambda$CDM cosmology~\cite{Ade:2015xua}.
The excitation amplitudes $A_{\l m}$, which depend on the mass ratio $q$ of the
binary, are determined using the fitting formulae in~\cite{Borhanian:2019kxt} 
at $t=10 M$ after the merger. The phases $\phi_{\l m}$ of the modes are
distributed uniformly in $\phi_{\l m} \in [0,2 \pi)$, in contrast to previous work in the
literature where both phases were fixed for simplicity
~\cite{PhysRevD.76.104044, Bhagwat:2019bwv, Gossan:2011ha}.

The BBH parameters for each candidate are drawn randomly from their
respective distributions to generate two-mode ringdown signals with
the dominant $\lm=(2,2)$ mode and either the $\lm=(3,3)$ or the
$\lm=(4,4)$ subdominant mode.  We consider a three-detector LIGO
network consisting of the observatories in Hanford (H1), Livingston
(L1) and India (I1). We use three different sensitivities for these
detectors~\cite{sensitivity_curves}: Advanced LIGO design sensitivity
(Adv. LIGO), A+ and Voyager.  We do not consider here the complete
third generation detectors, which include the Einstein Telescope
\cite{Sathyaprakash:2012jk,Sathyaprakash:2011bh,Punturo:2010zz} and
Cosmic Explorer \cite{Reitze:2019iox}, or the space based LISA mission
\cite{Danzmann:2003tv}, since this would take us beyond the 10-year
timeframe.

\begin{figure}
\centering
\includegraphics[width=\columnwidth]{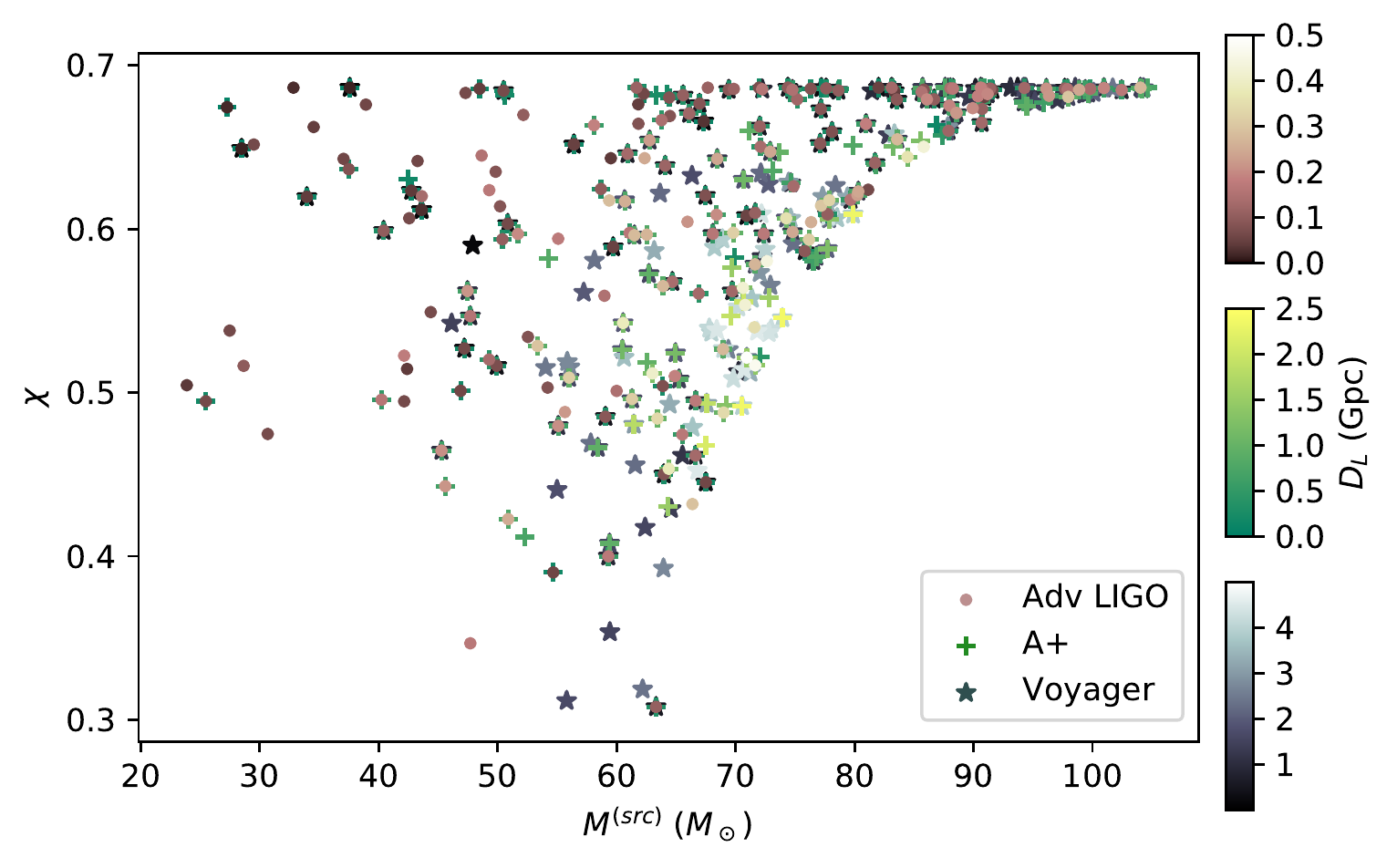}
\caption{\label{fig:population} Source frame mass, $M^{(\text{src})}$, 
and spin, $\chi$, of the BHs with optimal SNR $\rho_c \geq 2.5$ in the 
subdominant mode (either the $(3,3)$ or the $(4,4)$ mode), obtained 
using the observational population models of~\cite{LIGOScientific:2018jsj}.
The colors represent the luminosity distance of the source, where the 
maximum allowed distance was $D_L^{\text{(max)}} = \{1, 3, 5\}$ Gpc 
for Adv. LIGO, A+ and Voyager, respectively}
\end{figure}

For each candidate, we calculate the optimal SNR of the subdominant
mode in each detector,
$\rho_{\text{det}} = \sqrt{\langle h, h \rangle}$, where $h$ is the
ringdown signal of the subdominant mode projected into the detector
(see Eq.~\eqref{eq:h_ringdown}).
To avoid a large number of sources with no measurable subdominant mode, 
we reject candidates with combined optimal SNR
$ \rho_c = \sqrt{\sum_{\text{det}} \rho_{\text{det}}^2} < 2.5$ in the
subdominant mode. For the same reason, the maximum $D_L$ considered is
limited to different values for different sensitivities, namely
$D_L^{\text{(max)}} = \{1, 3, 5\}$ Gpc for Adv. LIGO, A+ and Voyager, respectively.  
The number of draws required to find a sample population of 100 signals
with $\rho_c \geq 2.5$ in the subdominant mode yields the fraction of
interesting candidates out of all BBH
signals. Figure~\ref{fig:population} shows the resulting populations
for each detector network considered.

\section{Analysis and results}
\label{sec:results}

\subsection{Rates of measurable subdominant modes}

We add the population of accepted candidate ringdown signals 
(shown in Fig.~\ref{fig:population}) into different Gaussian noise 
realizations colored with the PSD of the desired detector. 
To determine the measurability of the subdominant mode, 
we use the Kerr BH ringdown model and perform two separate 
Bayesian parameter estimation analyses using:
\begin{inparaenum}[($H_A$)]
\item templates with the fundamental $(2, 2)$ mode plus the 
	corresponding $\lm$ subdominant mode, and
\item templates with only the fundamental $(2, 2)$  mode.
\end{inparaenum}
The Bayes factor $\BF$ is then calculated as the ratio of the evidences 
for model $H_A$ versus model $H_B$. Those sources with $\BF > 3.2$ 
are further analyzed in the next section to determine the resolvability of 
the subdominant mode.

The parameters ($M$, $\chi$, $A_{\l m}$, $\phi_{\l m}$, $\iota$, $\psi$) are 
estimated from the data, which represents a set of 8 parameters in the two-mode 
ringdown $H_A$, and 6 parameters in the single-mode ringdown $H_B$.
The priors used in the parameter estimation analysis are uniform in all parameters: 
BH mass $M \in [10,200) M_\odot$, BH spin $\chi \in [-0.99,0.99)$, 
log-amplitude of the fundamental mode $\log_{10}(A_{22}) \in [-4,4)$, 
relative subdominant mode amplitude $\hat{A}_{\l m} = A_{\l m}/A_{220} \in [0,0.5)$,
ringdown phases $\phi_{\l m} \in [0, 2\pi)$, polarization angle $\psi \in [0, 2\pi)$, 
and inclination angle $\cos(\iota) \in [-1, 1)$. 
We fix the start time of the ringdown, the $\lm$ of the subdominant mode, 
the sky location and the distance to the source to the injected values.
While the start time of the ringdown is not uniquely defined in the literature, 
we do not explore the issue in this paper and assume
that this can be determined by other means~\cite{2018PhRvD..97l4069C, 
bhagwat2018choosing, Carullo:2019flw}. 
Further, we can safely assume that we have some knowledge from the inspiral 
part of the signal regarding the mass ratio of the binary 
to determine which is the loudest subdominant mode to look for. 
Since we are using a network of three detectors, the sky location should 
be relatively well known from the analysis of the full gravitational-wave signal.
Finally, while the distance might not be accurately measured, fixing this parameter 
to a wrong value will only affect the measurement
of the fundamental amplitude $A_{22}$ and not affect our conclusions. 

\begin{table*}[t]
\begin{tabular}{| c | c | c | c | c | c | c |}
\hline
 & \multicolumn{3}{c |}{$\beta_q = 0$} & \multicolumn{3}{c | }{$\beta_q = 6.7$} \\
\cline{2-7}
 & $\BF > 3.2$ & $\BF > 10$ & $\BF > 100$ & $\BF > 3.2$ & $\BF > 10$ & $\BF > 100$ \\
\hline
Adv. LIGO & $0.036^{+0.039}_{-0.019} $ & $0.028^{+0.031}_{-0.015}$ & $0.011^{+0.012}_{-0.006}$ & $0.008^{+0.009}_{-0.004}$ & $0.006^{+0.007}_{-0.003}$ & $0.003^{+0.003}_{-0.001}$ \\
\hline
A+ & $0.46^{+0.51}_{-0.25}$ & $0.28^{+0.31}_{-0.15}$ & $0.14^{+0.15}_{-0.07}$ & $0.08^{+0.09}_{-0.04}$ & $0.06^{+0.06}_{-0.03}$ & $0.03^{+0.03}_{-0.02}$ \\
\hline
Voyager & $2.63^{+2.89}_{-1.42}$ & $1.83^{+2.01}_{-0.99}$ & $0.89^{+0.97}_{-0.48}$ & $0.30^{+0.33}_{-0.16}$ & $0.21^{+0.24}_{-0.12}$ &  $0.11^{+0.12}_{-0.06}$ \\
\hline
\end{tabular}
\caption{Rates of BBH ringdown signals per year $(\text{yr}^{-1})$ with a detectable subdominant $(3,3)$ or $(4,4)$ mode for a population with uniform mass-ratio distribution ($\beta_q = 0$) and for a population with $\beta_q = 6.7$. The Bayes factors in each column indicate substantial support ($\BF > 3.2$), strong support ($\BF > 10$), and decisive support ($\BF > 100$) for the presence of a second mode. }
	\label{table:BFs}
\end{table*}

We calculate the rate of ringdown events with detectable subdominant mode in each
detector network based on the BBH merger rate density given in~\cite{LIGOScientific:2018jsj}  
($R = 53.2^{+58.5}_{-28.8}\text{ Gpc}^{-3}\text{ yr}^{-1}$) and the co-moving volume up to
$D_L^{\text{(max)}}$ for each detector network. Table \ref{table:BFs} lists the rate 
of events  per year with substantial ($\BF > 3.2$), strong ($\BF > 10$), and decisive ($\BF > 100$) 
support for the presence of a subdominant mode. These rates are the combination of 
both the $(3,3)$ and the $(4,4)$ modes. While we have made the simplifying assumption 
that only one subdominant mode will be measurable, some of the considered BBH systems 
might have two subdominant modes with SNR $\rho_c \geq 2.5$. However, studying the 
performance of a three-mode ringdown Bayesian analysis is beyond the scope of this paper.

\subsection{Resolvable subdominant modes for testing GR}

In the presence of two measurable ringdown modes, resolvability of the
$\Omega_{\l m}$ frequencies allows for BH spectroscopy tests. However,
QNMs of rotating BHs in modified theories of gravity have not been
calculated~\cite{Berti:2018vdi}, and Kerr-like exotic compact objects
can have the same or similar QNM spectrum as Kerr
BHs~\cite{Cardoso:2019rvt}. While it might be challenging to disprove
all BH alternatives, accurate measurements of the QNM spectrum will be
crucial to constrain deviations from GR (see however
\cite{Maselli:2019mjd} for possible ways of parameterizing frequencies
and damping times accounting for deviations from GR).  It has been
shown for non-rotating alternative BH models that GR deviations are
more significant in the QNM frequencies than in the damping
times~\cite{Moulin:2019ekf}. Hence, we focus here on constraining
deviations from the subdominant mode's frequency.

We consider those ringdown events with $\BF > 3.2$ in the previous 
section and perform the same parameter estimation analysis,
now using the agnostic model defined in Sec.~\ref{sec:ringdown} to 
estimate the ringdown $\Omega_{\l m}$ frequencies of the two QNMs.
Hence, 10 parameters ($\omega_{\l m}$, $\tau_{\l m}$, $A_{\l m}$, 
$\phi_{\l m}$, $\iota$, $\psi$) are now estimated from the data.
The priors are uniform in the frequencies 
$ f_{\l m} =  \omega_{\l m} / {2 \pi} \in [50, 1024)$ Hz 
and damping times $\tau_{\l mn} \in [0.45, 30)$ ms,
excluding parameters that yield masses and spins outside of the ranges 
used in the previous section with the Kerr model.
The amplitudes of the $\lm$ modes have different orders of magnitude, 
because of the missing factor $M/D_L$ when dropping the Kerr assumption. 
Hence, the prior in log-amplitude of the fundamental mode is now 
$\log_{10}(A_{22}) \in [-25,-17)$. The priors in the remaining parameters are 
the same as in the previous section. Finally, we apply an additional set of 
constraints on the subdominant frequency and damping time to be within 
$\pm 25\%$ of the GR expectation.

\begin{figure}[b]
\centering
\includegraphics[width=\columnwidth]{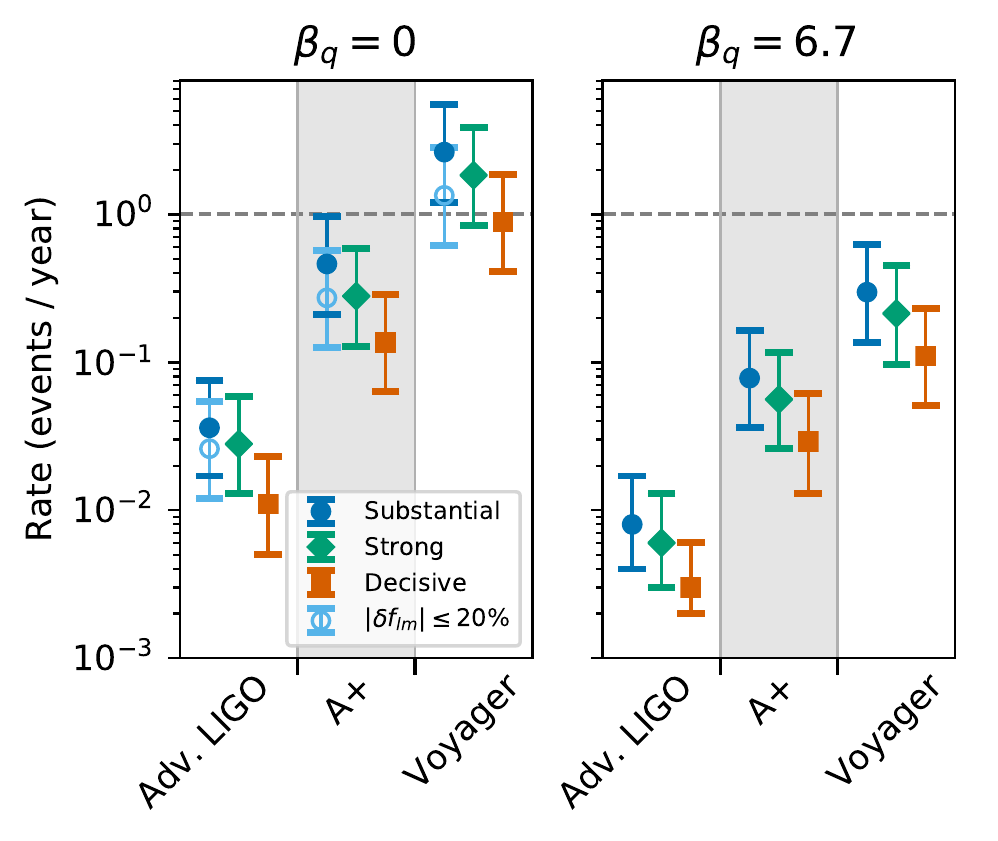}
\caption{Expected rates of BBH mergers for which two ringdown modes can be
observed and resolved. We consider two population models corresponding to
$\beta_q = \{0,6.7\}$. Shown are the rate of events that have ``substantial",
``strong", and ``decisive" Bayesian evidence (using the nomenclature of
Ref.~\cite{Kass:1995loi}) for a two-mode Kerr hypothesis relative to a
single-mode hypothesis (filled circles, diamonds, and squares, respectively).
Of the events that have substantial evidence, we perform a followup analysis in
which the frequency and damping time of the subdominant mode is allowed to
deviate from the expected GR value. The rate of events for which the deviation
from GR of the subdominant frequency $|\delta f_{lm}|$ is constrained to be
$\leq 20\%$ is given by the open circles.}
\label{fig:rates}
\end{figure}

Using the fitting formulae in~\cite{Berti:2005ys}, we can compare the mass and 
spin measurement obtained from the $(2, 2)$ parameters and from the subdominant 
$\lm$ parameters. Furthermore, based on the measurement of the $(2, 2)$ mode, 
we can infer the measured deviation on the frequency of the subdominant $\lm$ 
mode, $\delta f_{\l m}$. Table~\ref{table:deltas} lists the rates of BBH ringdown 
signals per year that constrain GR within $\delta f_{\l m} \pm 20\%$ at the $90\%$ 
credible level. The results are summarized in Fig.~\ref{fig:rates}.  

\begin{table}[h]
	\begin{tabular}{| c | c | }
		\hline
		Network & $\delta f_{\l m} \leq \pm 20\%$ \\
		\hline
		Adv. LIGO & $0.026^{+0.028}_{-0.014}$  \\
		\hline
		A+ & $0.27^{+0.30}_{-0.15}$  \\
		\hline
		Voyager & $1.34^{+1.47}_{-0.73}$ \\
		\hline
	\end{tabular}
	\caption{Rates of BBH ringdown signals per year $(\text{yr}^{-1})$ with strong support for the presence of a second mode ($\BF > 3.2$) where deviations of the GR frequencies are constrained to within $\delta f_{\l m} \leq \pm 20\%$ at the $90\%$ credible level. We only show the rates for the population with uniform mass-ratio distribution ($\beta_q = 0$), since we know from the previous section that rates for a population with $\beta_q = 6.7$ will be lower.}
	\label{table:deltas}
\end{table}

\section{Conclusions}
\label{sec:conclusions}

In this paper we have applied for the first time the full Bayesian inference
framework to a population of BH ringdowns derived from the observational
population models published by the LIGO Scientific and Virgo Collaborations.
Furthermore, we have allowed for completely variable ringdown phases,
inclination angles, polarization angles and sky locations, contrary to 
previous works that have fixed one or more of these parameters for 
simplicity~\cite{PhysRevD.76.104044, Bhagwat:2019bwv, Gossan:2011ha}. 

Within the Bayesian model selection framework, future generations
of LIGO detectors will likely deliver measurable subdominant QNM 
modes from BBH mergers over the next decade. However, resolvability 
of the subdominant frequencies is technically challenging, and accurate 
tests of the no-hair theorem might only be possible in very few cases.
These results are in agreement with previously published 
works~\cite{Berti:2016lat, Baibhav:2018rfk}, where the ringdown SNR 
was used to determine the measurability and resolvability of QNMs.

Merger population models from gravitational-wave observations are still largely
uncertain. The third observing run of Advanced LIGO and Virgo might be
uncovering a new population of NSBH and other previously unobserved types of
mergers, which could boost the rates of measurable and resolvable subdominant
modes. Hence, the rates obtained in this work might turn out to be pessimistic
as more gravitational-wave detections are made available.

\section{Acknowledgements}

We are thankful to Swetha Bhagwat, Duncan Brown, Evan Goetz, Scott
Hughes, Alexander H. Nitz, Paolo Pani and Frans Pretorius for useful
comments and discussions. We are especially grateful to Ssohrab
Borhanian and Bangalore Sathyaprakash for providing us with the fits
for the ringdown mode amplitudes in \cite{Borhanian:2019kxt}.  MC
acknowledges support from NSF grant PHY-1607449, the Simons
Foundation, and the Canadian Institute For Advanced Research (CIFAR).
Computations have been performed on the Atlas cluster of the Albert
Einstein Institute (Hannover).

\section{Appendix}
\label{app:results}

Results for the injection with the largest Bayes factor in the (3, 3) population using the Voyager sensitivity.
This BBH is located at a distance $D_L \simeq 250$ Mpc.

\begin{figure}
\centering
\includegraphics[width=\columnwidth]{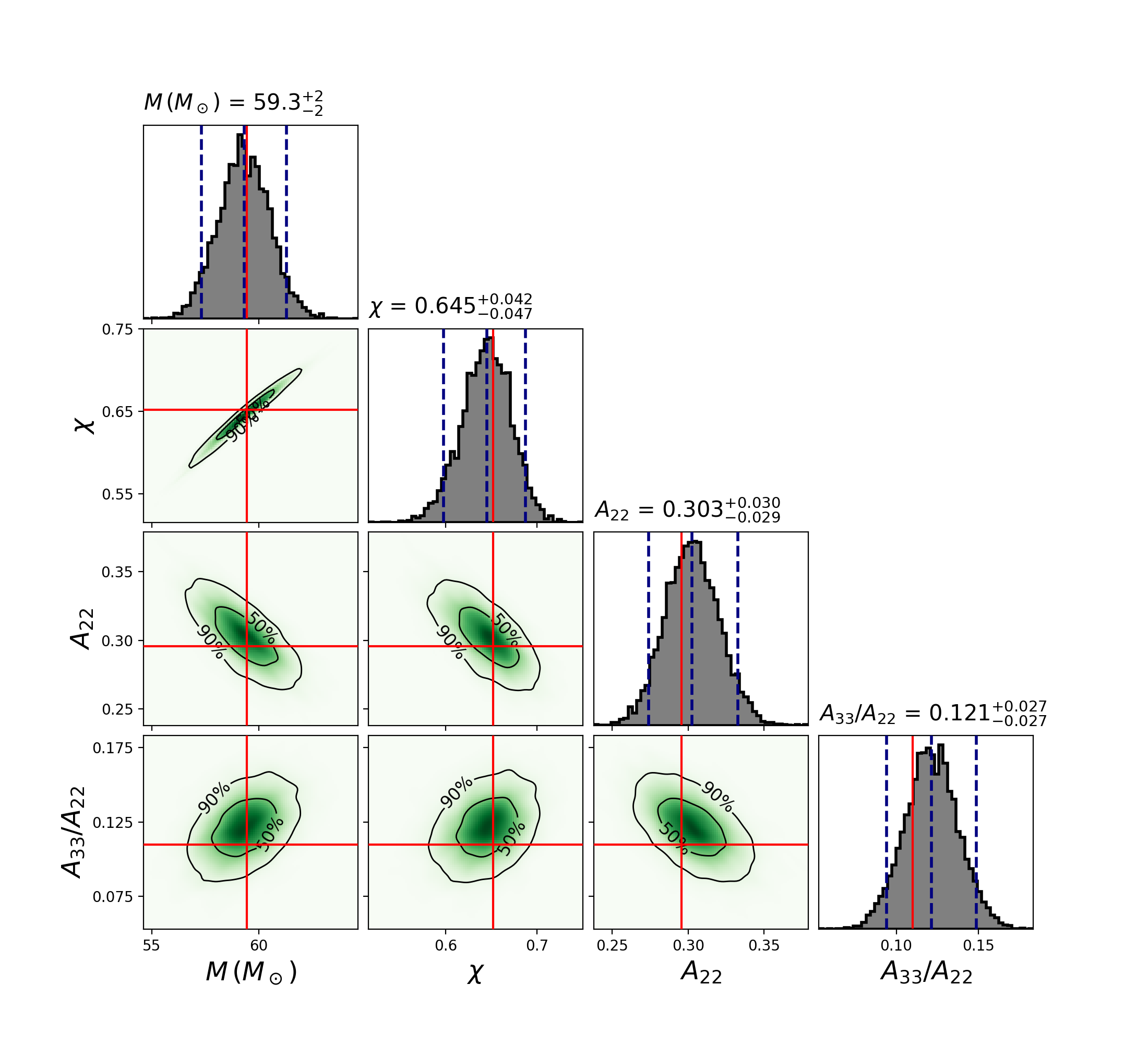}
\caption{\label{fig:kerr_posterior} Posterior distributions from the analysis with a Kerr ringdown. The parameters of interest are the BH mass $M$, BH spin $\chi$,
amplitude of the $(2, 2)$ mode, $A_{22}$, and amplitude ratio of the $(3, 3)$ mode, $A_{33} / A_{22}$. The red crosses
indicate the injected parameters, and the dashed lines in the histograms correspond to the median value and the 90\% credible level.}
\end{figure}

\begin{figure}
\centering
\includegraphics[width=\columnwidth]{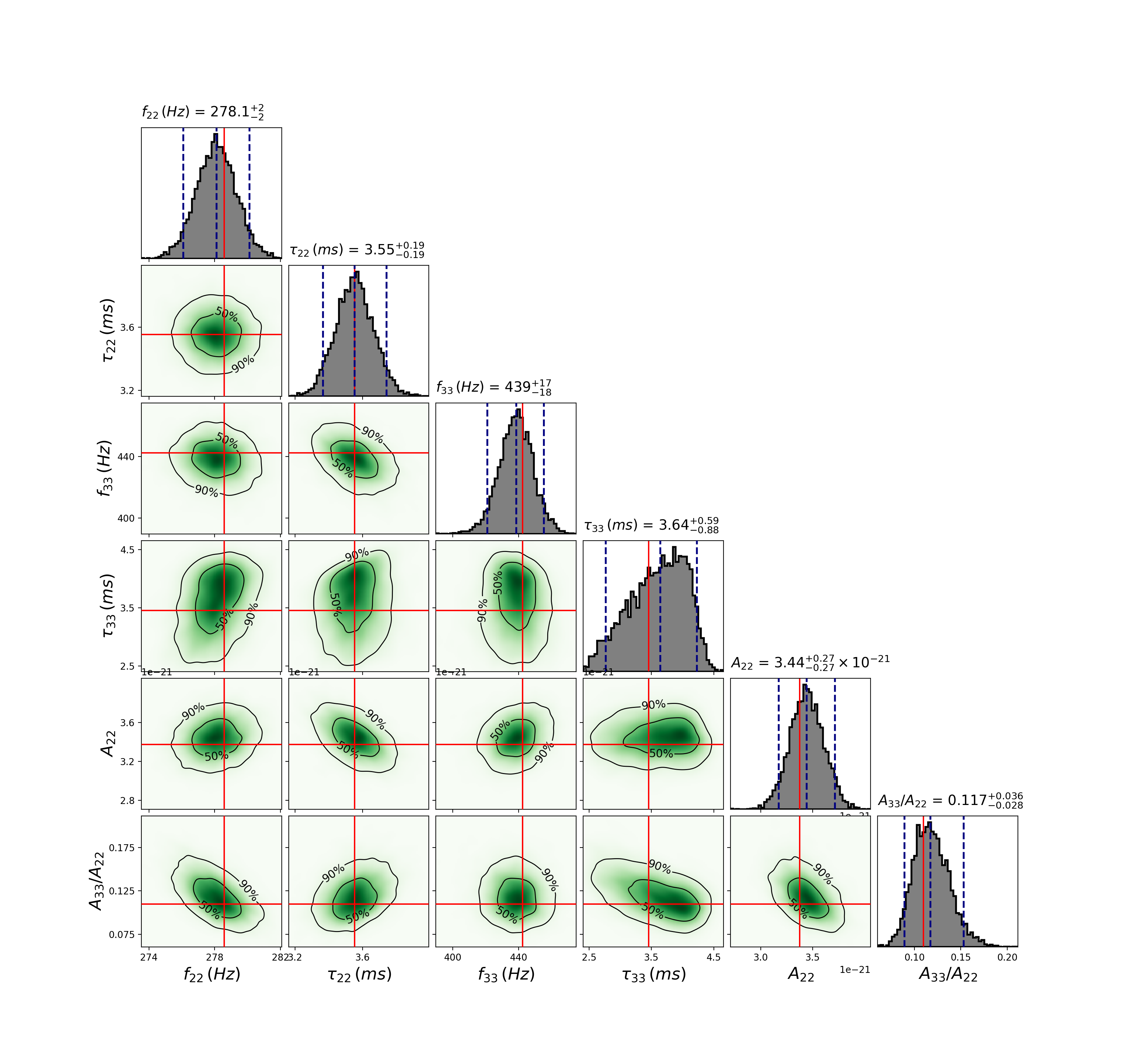}
\caption{\label{fig:freq-tau_posterior} Posterior distributions from the analysis with an agnostic ringdown. The parameters of interest are the 
ringdown frequencies, $f_{lm}$, and damping times, $\tau_{lm}$, the amplitude of the $(2, 2)$ mode, $A_{22}$, and the amplitude ratio of the $(3, 3)$ mode, $A_{33} / A_{22}$.
$A_{22}$ has a different value because of the missing $M/D_L$ factor in the approximant. The red crosses
indicate the injected parameters, and the dashed lines in the histograms correspond to the median value and the 90\% credible level.}
\end{figure}

\bibliographystyle{ieeetr}
\bibliography{ringdown}

\end{document}